# Impact of Cr-O hyrbidization in ACrO$_3$ (A=La, Y): A Theoretical Investigation

Jeel Swami[1] Ambesh Dixit[2] and Brajesh Tiwari[1]*

[1]Department of Basic Sciences, Institute of Infrastructure, Technology Research and Management Ahmedabad, 380026 India

[2]Department of Physics, Indian Institute of Technology Jodhpur, Jodhpur, India-342037

Corresponding author email: brajeshtiwari@iitram.ac.in

**Abstract**:

Electronic properties of spin polarized antiferromagnetic ACrO$_3$ (A = La, Y) are explored with Hubbard Model using Density Functional Theory (DFT). These two isostructural systems are investigated using the different Hubbard energy and analyzed the hybridization of chromium 3d orbitals and oxygen 2p orbitals and the change in energy band gaps against the Hubbard energy. The bond length and bond angle affect significantly the orbital contributions of Cr-3d and O-2p electrons for both the system. We noticed that the Cr-O hybridization affects the orbital degeneracy and is substantiated with partial density of states. These results emphasize the contribution of Hubbard energy in correlated electron systems.

# 1. INTRODUCTION

ABO$_3$ type perovskite crystal structure with B as a transition metal ion has been immensely popular in the research community over the past several decades [1]. Rare earth ferrites ReFeO$_3$ and Rare earth orthochromites ReCrO$_3$ (Re= La-Lu, Y) exhibit distinctive structural, electronic, magnetic, and optical properties like magnetocaloric effect, magnetic exchange bias, photocatalysis, anisotropic magnetostriction, magnetoelectric-multiferroicity, and relaxor dielectric behavior [2–11]. These materials exhibit antiferromagnetic ordering together with Dzyaloshinsky-Moriya type interaction, giving rise to magnetoelectric and multiferroic properties. [12,13]. Rare earth orthochromite shows orthorhombic GdFeO$_3$-type of distorted perovskite crystal structure with Pnma space group [14,15]. Their exciting properties, like ferroelectricity, complex magnetic structure, colossal magnetoresistance (CMR), metal-insulator transitions, etc., are the manifestation of tilting and distortion of the respective octahedra [1,16–18]. These materials are used in solid oxide fuel cells, photovoltaic solar cells, and electrochemical and photochemical water splitting because of their suitable optoelectronic properties [19–23]. These findings have sparked a lot of curiosity to study different properties of rare earth chromites.

For the ideal perovskite ABO$_3$, structure has cubic symmetry with bond angle B-O-B. Distorted perovksite structure gives them certain unusual physical properties. Further, ACrO$_3$ distorted perovskites have attracted their attention because of their diverse properties with simple structures [24–27]. The size of A ion can distort orthorhombic crystal structure by tilting or rotating $CrO_6$ octahedra [28]. Tilting or rotation of octahedra reduces Cr-O-Cr bond angle and thus, giving rise to canted antiferromagnetism in such systems. Y (yttrium) is a transition metal and is explored with rare earth elements (La-Lu) because of their similar ionic radii. Here, we considered transition metal (Y) and rare earth (La) based chromites having the same crystal structure and symmetry to understand the significant change in their physical properties at a microscopic level. If we took two isostructural systems, ACrO$_3$ (A=La, Y) we expect that this Cr-O-Cr bond angle distortion will play a major role in Cr-O hybridization and this will later affect the crystal field splitting of Cr(3d) orbitals. Uzma et al. concluded that d-orbitals of transition metal is accountable for the unique electronic properties of the compounds [29].

This work is motivated to investigate the Cr-O hybridization under the structural distortion happening in two isostructural system. We expect that Cr-O-Cr bond angle distortion plays a crucial role when two different atoms with different ionic radii is substituted on A-site.

Moreover, for the correlated system we have to apply on-site column interaction on Cr-site. Hence, it will be interesting to analyze on-site coulomb interaction on correlated system ACrO$_3$ (A=La, Y) and its effect of on Cr-O hybridization. The significance of this work is effect of Hubbard energy on Cr-O hybridization and performing a comparative first principle study analyse structural, electronic and magnetic properties of LaCrO$_3$ and YCrO$_3$. Moreover, Cr-O hybridization affects the crystal field splitting of ACrO$_3$.

LaCrO$_3$ shows structural phase transition, at higher temperatures, from orthorhombic to rhombohedral at 526 K and rhombohedral to cubic at 1300 K, respectively. Geller et al. showed that it changes to an ideal cubic structure at or above 1900 K [30–33]. Thus, LaCrO$_3$ can exhibit orthorhombic, rhombohedral, and cubic phases at different temperatures. YCrO$_3$ shows antiferromagnetic transition below 140 K and ferroelectric ordering at ~ 473 K in the same phase without any structural transition [34]. LaCrO$_3$ is G-type AFM (at 290 K), whereas YCrO$_3$ is an antiferromagnet with weak ferromagnetism (T$_N$~140 K) [8,31,35]. Yogesh et al. reported YCrO$_3$ as a biferroic material with G-type AFM below T$_N$ ~ 142 K and ferroelectric transition T$_C$ ~ 473 K, where weak polarization and a dielectric anomaly were observed [34,36]. Vidhya et al. studied the electronic properties of YCrO$_3$ and showed that the G-type AFM structure is stable for YCrO$_3$ [37]. Thus, there are various reports on both LaCrO$_3$ and YCrO$_3$ independently, where properties differ significantly, even being isostructural systems. However, to the author's best, there is no report comparing these systems systematically to understand the microscopic changes leading to different characteristics.

We used the density functional theory (DFT) to study the structural, electronic and magnetic properties of LaCrO$_3$ and YCrO$_3$ with varying Hubbard energy for capturing the contribution from Cr-3d, and O-2p orbitals near Fermi level, using their weighted electronic band structure and density of states (DOS). In this paper, impact of Hubbard energy on ACrO$_3$ is investigated with two approaches GGA and GGA+U.

**COMPUTATIONAL METHOD**

The DFT calculations are performed using Generalised Gradient Approximation (GGA) parameterized by Perdew, Burke, and Ernzerhof as implemented in Quantum Espresso (QE) [38], and projected augmented wave (PAW) pseudopotentials are used [39]. GGA with Density functional theory (DFT) and GGA augmented with the Hubbard U method are implemented in this work [40]. For strongly correlated materials, the standard DFT methods like local-density approximation (LDA) and GGA fail because of not including the on-site Coulomb interaction. This limitation of standard DFT can be taken care of by introducing DFT+U method [41,42]. The DFT + U method, which combines GGA calculations with an additional orbital-dependent interaction and it takes into account for orbitals of the atomic type that are highly localised on the same site. DFT+U method is used to better describe strongly correlated systems and to correct the errors introduced by the standard DFT for these systems. The DFT + U method, which explicitly accounts for the Coulomb interaction between electrons at a single site in the total energy functional is equivalent to the Hubbard model U and reduce the effect of residual self-interaction. Hubbard parameter consists of the on-site Coulomb interaction parameter U and exchange interaction parameter J as

$$E^{DFT+U+J} = E^{DFT} + \frac{1}{2}\sum_{I,\sigma}\sum_{nl}(U_{nl}^{I} - J_{nl}^{I})\left[Tr\left(\boldsymbol{n}_{nl}^{I,\sigma}(1 - \boldsymbol{n}_{nl}^{I,\sigma})\right)\right] + \frac{1}{2}\sum_{I,\sigma}\sum_{nl}J_{nl}^{I}[Tr(\boldsymbol{n}_{nl}^{I,\sigma}\boldsymbol{n}_{nl}^{I,-\sigma})]$$

Where U and J are spherically averaged matrix element of on-site Coulomb interactions, and $\boldsymbol{n}$ is on-site nl-orbital (here, 3d) occupation matrix produced by projecting of wave function onto 3d atomic-like states and subscript nl represent d-orbital index whereas spin is denoted by $\sigma$. The above equation demonstrates the on-site occupation matrix with orbital and spin representation. Here, we only consider Hubbard (on-site Coulomb interaction) term U and we did not include J (exchange interaction) in our calculation. Hence, the equation $U_{eff} = U$. Hubbard energy we applied to the system will become the effective U. The electronic and magnetic properties of the system calculated using standard DFT and then applied Hubbard U as a correction parameter. The electronic self-consistency calculations are performed using 6×6×6 Γ-centered k-point grid in the primitive unit cell. Plane-wave energy cutoff of $LaCrO_3$ and $YCrO_3$ are optimized and set at 480 eV and 440 eV, respectively. The valence configuration for La: $5d^1 6s^2$, Y: $4d^1 5s^2$, Cr: $3d^5 4s^1$, and $2s^2 2p^4$ for O are used for this calculation. Spin-polarized density functional theory-based calculations are carried out to study their structural, electronic and magnetic properties.

## 2.   RESULTS AND DISCUSSION

**Structural Properties**

ACrO$_3$ (A=La, Y) has an orthorhombic perovskite structure with Pnma (#62) space group. The calculation is carried out for G-type antiferromagnetic-structure [21]. ACrO$_3$ crystal structure is shown in Figure 1(a), where lattice parameters $b > a > c$ and angles $\alpha = \beta = \gamma = 90°$. Cr ions are surrounded by six O atoms and make $CrO_6$ octahedra, as shown in Figure 1(b). Figure 1(b), the nomenclature of O is given in $CrO_6$ octahedra, O1-4 are non-axial oxygen whereas O11 and O12 are axial oxygen. The lattice parameters 'a' is optimized and total energy as a function of lattice parameter is plotted in Figure 1 (d). Relaxed fractional atomic position and site symmetry of non-axial O (O1-4), axial O (O11/O12), Cr and La/Y of ACrO$_3$ are listed in supplementary S1. A (A=La, Y) having (x, 0.25, y), Cr has (0.0, 0.0, 0.5), O11-12 has (x, 0.75, z) and O1-4 has (x, y, z) Wyckoff position. The optimized lattice parameters for LaCrO$_3$ are a=5.6093 Å, b=7.8846 Å, c=5.5646 Å and for YCrO$_3$ are a=5.5564 Å, b=7.8102 Å, c=5.5124 Å and they are in good agreement with earlier work as shown in Table 1 indicating the reliability of our calculation (detailed comparison of optimized lattice parameters with reported experimental and computed work are listed in Supplementary S2). [7,8,28,43–47]. GGA-PBE method is overestimates the lattice parameters of the system. Cr-O bond lengths and Cr-O-Cr bond angles of ACrO$_3$ are listed in Table 2. It shows that bond lengths (in pair) Cr-O1 and Cr-O2; Cr-O3 and Cr-O4; Cr-O11 and Cr-O12 are the same. Moreover, Cr in ACrO$_3$ is in +3 oxidation state, which is John-teller inactive [48,49]. It implies that there is no further distortion in $CrO_6$ octahedra and Cr is intact on its centrosymmetric position in both LaCrO$_3$ and YCrO$_3$. The bond angle between axial Cr-O-Cr should be 180° for ideal perovskite but here the angle is reduced from 180° in both the cases, suggesting $CrO_6$ octahedra tilting. This tilting of octahedra affects superexchange coupling and causes changes in the magnetic ordering temperature [50,51]. The antiferromagnetic ordering temperature for LaCrO$_3$ is 290 K and for YCrO$_3$ is 140 K, as observed and reported in our earlier experimental work [4,8]. The axial bond angle Cr-O-Cr is 159.512° and 145.430° for LaCrO$_3$ and YCrO$_3$, respectively, comparable to the experimental reports [4,8]. The bond angle for YCrO$_3$ is far from 180°, which promotes sizeable octahedral distortion compared to that of LaCrO$_3$. LaCrO$_3$ and YCrO$_3$ have the same structure but different bond angles, affecting their properties. Further, check the statbility of the system, tolerance factor is calculated [52]. Tolerance factor for perovskites of the ABO$_3$ is defined as,

$$t = \frac{r_A + r_O}{\sqrt{2}(r_B + r_O)}$$

From this tolerance factor values of LaCrO$_3$ and YCrO$_3$ are 0.8997 and 0.8499, respectively which suggest the orthorhombic crystal structure.

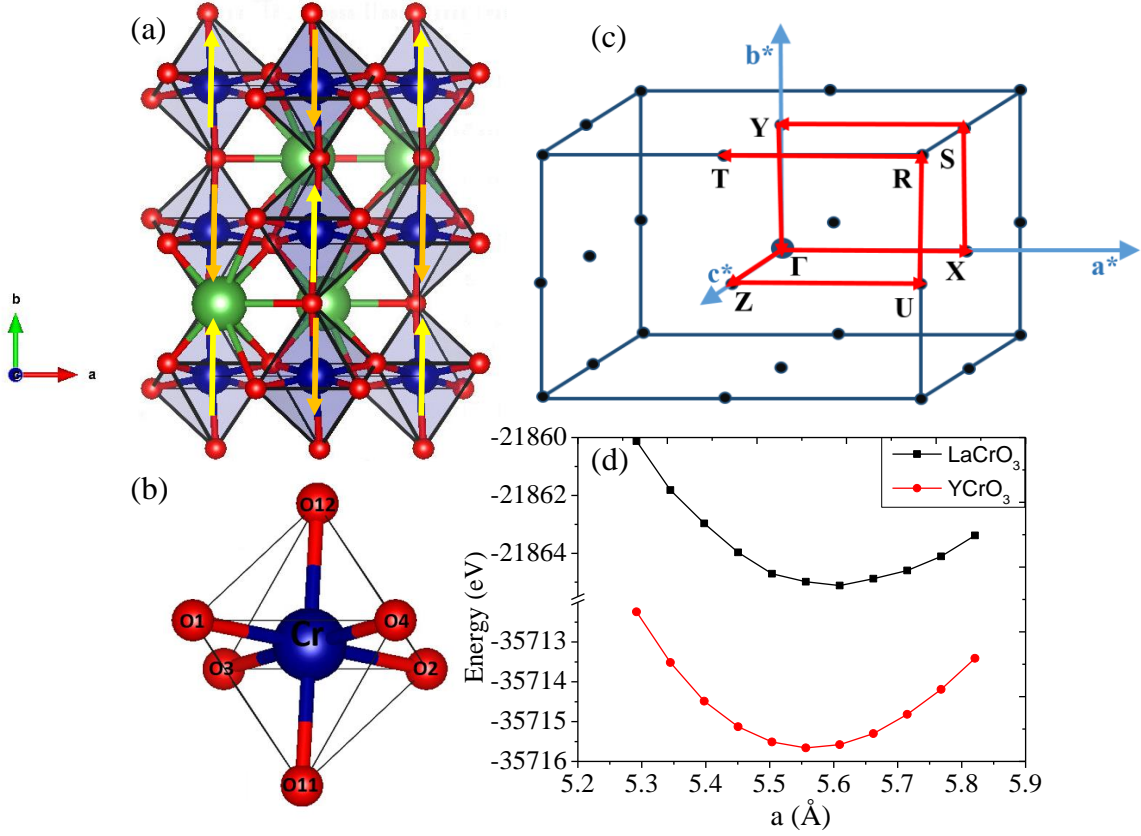

FIG. 1(a). Crystal structure of orthorhombic ACrO$_3$. Atoms with green color represent A (La/Y) atoms, blue represents Cr, and red represents O atoms. The black lines show the octahedral in the system. The yellow arrow shows the spin direction of Cr atoms to make G-type ACrO$_3$ antiferromagnetic system. (b) $CrO_6$ Octahedral where $Cr^{+3}$ surrounded by octahedral oxygen cages, (c) Brillion zone path in a primitive unit cell. (d) Energy minimization curve where total energy of ACrO$_3$ (A=La, Y) as a function of the lattice parameter.

TABLE 1: Lattice parameters of ACrO$_3$ (A=La, Y).

Lattice parameters (Å)

| | | | | | |
|---|---|---|---|---|---|
| LaCrO$_3$ | a=5.6117 | a=5.494 | a=5.5345 | a=5.4179 | a=5.6093 |
| | b=7.8880 | b=7.777 | b= 7.7892 | b=7.759 | b=7.8846 |
| | c=5.5673 | c=5.520 | c=5.4957 | c=5.516 | c=5.5649 [present work] |
| | [43] | [28] | [45] | [7,8] | |
| YCrO$_3$ | a=5.6147<br>b=7.6674<br>c=5.3172<br>[43] | a= 5.5157<br>b= 7.5301<br>c= 5.2409<br>[44] | a=5.5239<br>b=7.5363<br>c=5.2438<br>[46] | a=5.520<br>b=7.536<br>c=5.255<br>[28] | a=5.5564<br>b=7.8102<br>c=5.5124 [present work] |

TABLE 2: Cr-O bond length and Cr-O-Cr bond angle in ACrO$_3$ (A= La, Y) system.

| System | Bond length (Å) | | | Bond angle (°) | |
|---|---|---|---|---|---|
| | Axial | Non-axial | | axial | Non-axial |
| | **Cr-O11/12** | **Cr-O1/2** | **Cr-O3/4** | **Cr-O11-Cr** | **Cr-O1-Cr** |
| LaCrO$_3$ | 2.0254<br>1.9749 | 2.0299<br>1.9733 | 2.0261<br>1.9733 | 159.512<br>161.36 | 159.249<br>159.07 [30] |
| YCrO$_3$ | 2.0449<br>1.971 | 2.0384<br>1.984 | 2.0390<br>1.993 | 145.430<br>145.53 | 147.396<br>146.22 [53] |

**Electronic Properties**

For the electronic band structure calculation, the Γ-X-S-Y-Γ-Z-U-R-T path is chosen, as shown in Figure 1(c). Energies are taken with respect to the Fermi level for band structure, and density of states, where negative energy values refer to valence bands and positive energy values refer to the conduction bands. Here, near the Fermi level ($E_F$), the contribution of Lanthanum and Yttrium is negligible; hence the contribution of 3d orbitals of Cr and 2p orbitals of O is shown with a total density of states. The spin-polarized electronic band structure of ACrO$_3$ with high symmetry points in the Brillouin zone is shown in Figure 2(a1 and b1). The band structure of ACrO$_3$ with different U are plotted in Figure 2 (a2-b5) along with their density of state (TDOS).

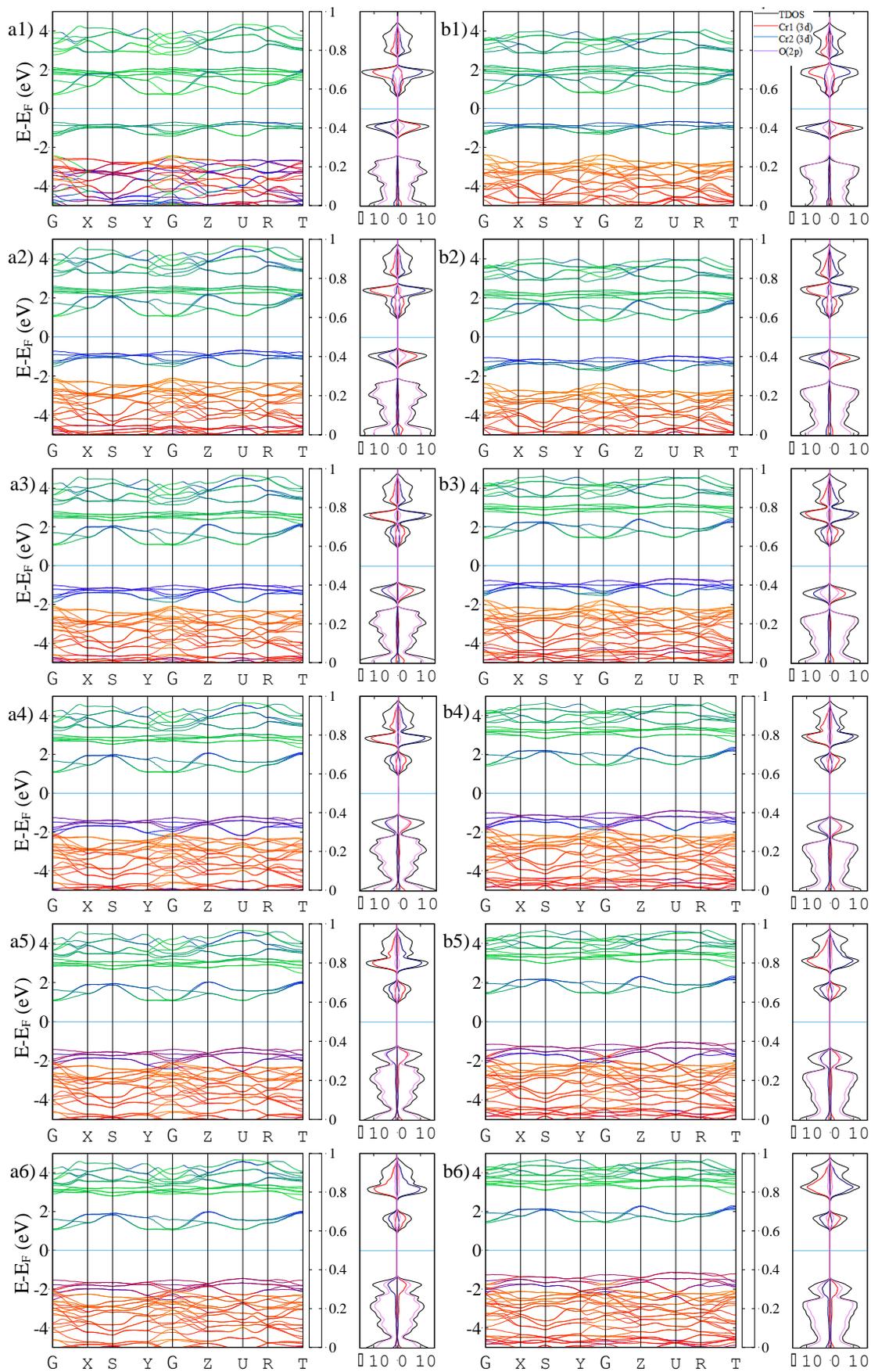

Figure 2: Band diagram with total Density of state for (a1) LaCrO$_3$, (b1) YCrO$_3$ with U = 0 eV, (a2) LaCrO$_3$, (b2) YCrO$_3$ with U = 1 eV, (a3) LaCrO$_3$, (b3) YCrO$_3$ with U = 2 eV, (a4) LaCrO$_3$, (b4) YCrO$_3$ with U = 3 eV, (a5) LaCrO$_3$, (b5) YCrO$_3$ with U = 3.7 eV, (a6) LaCrO$_3$, (b6) YCrO$_3$ with U = 4.3 eV. (Color line) In the color bar of electronic band diagram, '0' (green color) represent Cr contribution whereas '1' (orange color) represents O contribution. In density of state, black color represents total density of states (TDOS), red color represent contribution of Cr1 3d (spin up), blue color represent contribution of Cr2 3d (spin down) and purple color represent O 2p contribution.

We used corrected GGA + U calculations using the PBE technique to estimate the band value of the correlated system ACrO$_3$ in order to take into consideration the high electron correlations of the chromium 3d electrons. Between 0 and 4.3 eV, the Hubbard parameter U for Cr was changed. In Figure 2, the color bar shows the contribution of 3d orbitals of Cr and 2p orbitals of O as '0' and '1', respectively, i.e., green colored bands show the contribution of Cr 3d orbitals whereas orange colored bands show the contribution of O 2p orbitals. Here, we used weighted electronic band diagrams, which help to visualize the Cr-O hybridization happening in both isostructural systems. The nature of electronic band diagrams of LaCrO$_3$ and YCrO$_3$ seem quite similar, as can be seen in weighted electronic band diagrams in Figure 2(a1) and (b1), yet the contribution is not the same. We noticed YCrO$_3$ shows more O 2p contribution than that of LaCrO$_3$. We explained the hybridization happening between 3d orbitals of Cr and 2p orbitals of O, because of the different ionic radii of La and Y in ACrO$_3$. $Cr^{+3}$ ions have partially filled d orbitals with 0.615 Å ionic radii whereas $La^{+3}$ and $Y^{+3}$ have larger ionic radii than compare $Cr^{+3}$ and Shannon ionic radii of $La^{+3}$ is 1.032 Å and $Y^{+3}$ is 0.9 Å in VI coordination [54]. La and Y are placed on A site of ACrO$_3$, and significant difference is created in both systems due to their different ionic radii. Ionic radii of $Y^{+3}$ is less than $La^{+3}$, inducing structural instability. YCrO$_3$ is more distorted from ideal perovskite compared to LaCrO$_3$. The Cr-O-Cr bond angle difference can also explain it. In YCrO$_3$, the Cr-O-Cr bond angle is 145.43°, whereas in LaCrO$_3$, it is 159.51°, explaining more contribution of O 2p in YCrO$_3$, leading to more Cr-O hybridization. The changes are also visible in DOS from the right panel of Figure 2 (a, b). In the electronic band diagram, Cr contribution dominates (green) in bands above Fermi level and bands just below Fermi level. It is observed from the density of states that when there is no Hubbard energy applied, the bands near Fermi level and bands between 3 to 4 eV, majorly Cr is contributing together with a small contribution of O as well. O 2p orbitals have major contribution in between -2 to -5 eV energy bands.

As U increases on the Cr site, contributions of the bands are also changing, as shown in Figure 2. The contributions from Cr and O are changing for $LaCrO_3$ (a1-a6) and $YCrO_3$ (b1-b6), along with the change in energy gaps. In Figure 2 (a3, b3) where U = 2 eV, five bands of Cr(3d) just above Fermi level began to split into two parts, bands only contain Cr 3d contribution and Cr(3d) O(2p) hybridized bands with both Cr 3d and O 2p contributions. This splitting of bands can be seen clearly in their respective DOS. However, bands just below the Fermi level with major Cr(3p) contribution move towards lower energy and merge with lower energy bands of O(2p) when U=3.7 eV, as shown in Figure 2(a4, b4). Merging of bands below the Fermi level signifies the Cr-O hybridization. In Figure 2(a5, b5) and (a6, b6), the bands above the Fermi level are splitted and merged with the higher energy bands. The Cr(3d)-O(2p) hybridization also increases with increasing Hubbard energy. When U = 0 eV, Cr 3d orbitals contribute to bands just below Fermi level and O 2p orbitals separated by a charge-transfer gap. When U = 4.3 eV, the charge transfer gap vanishes due to hybridizing Cr(3d)-O(2p) bands. This trend is observed in both the $ACrO_3$ system with U. In density of states, when U = 0 eV near Fermi level 3d orbitals of Cr are majorly contributing, and below Fermi level 2p orbitals of O are majorly contributing. As the U increases, the charge transfer gap (between O 2p orbitals and Cr 3d orbitals) vanishes, and contribution of O-2p orbitals will dominate more below Fermi level with the hybridization of Cr and O orbitals. We can also observe that the existence of optical and charge transfer gap which varnish with GGA+U approach.

Figure 3 (a) shows the schematic changes in $ACrO_3$ system without and with Hubbard energy. It is observed that the band gap is increasing with increasing Hubbard U. The band gap as a function of Hubbard U is plotted for both $LaCrO_3$ and $YCrO_3$ in Figure 3(b), showing increasing bandgap with Hubbard energy. Energy band gap values with Hubbard energy for $ACrO_3$ (A=La, Y) is listed in Supplementary S3. Experimental values of energy band gaps are in good agreement when Hubbard energy is kept at 4.3 eV for $LaCrO_3$ and $YCrO_3$. Similar first principle study of change in energy band gap with Hubbard energy was done by Mourad Rougab and Ahmed Gueddouh which also support the increment of energy band gap with increasing U [55]. Thus, Coulomb repulsion in these systems results in increasing the bandgap while using DFT+U for these systems. There are several methods to choose right Hubbard U value for particular system [55–58]. In GGA+U, Hubbard parameter U is an empirical correction parameter and it can be tailored from existing experimental results (like thermodynamic values, energy band gap values, charge localization and kinetic barriers). Different fitting parameter give different values of U. Here, we choose to fit experimental the lattice parameters and energy band gap of ACrO$_3$ by applying Hubbard parameter. Similarly, Ali et al. have investigated BiFeO$_3$ and BaTiO3 by choosing the Hubbard U value comparing with experimental band gap value [59]. Energy band gap value with Hubbard energy is listed in Supplementary S3 and band gap values are increasing with increasing Hubbard U as shown in Figure 4 (b). Here, the focus of the paper is to investigate the effect of Hubbard U on ACrO$_3$, rather than choosing right Hubbard value for the system. The bands are separated by three energy gaps (charge transfer gap, band gap, and gap above Fermi level) without U in these systems. Charge transfer gap separates Cr 3d and O 2p bands; band gap is separated by occupied and unoccupied bands and another energy gap is observed above the Fermi level. The charge-transfer gap vanishes with U and the bands below Fermi level have merged to lower the energy bands. The band gap increases with increasing U and the band above the Fermi level split into two parts: (i) one-part merge into the upper energy bands, and (ii) other bands will remain separated from the upper energy bands and reside above the Fermi level.

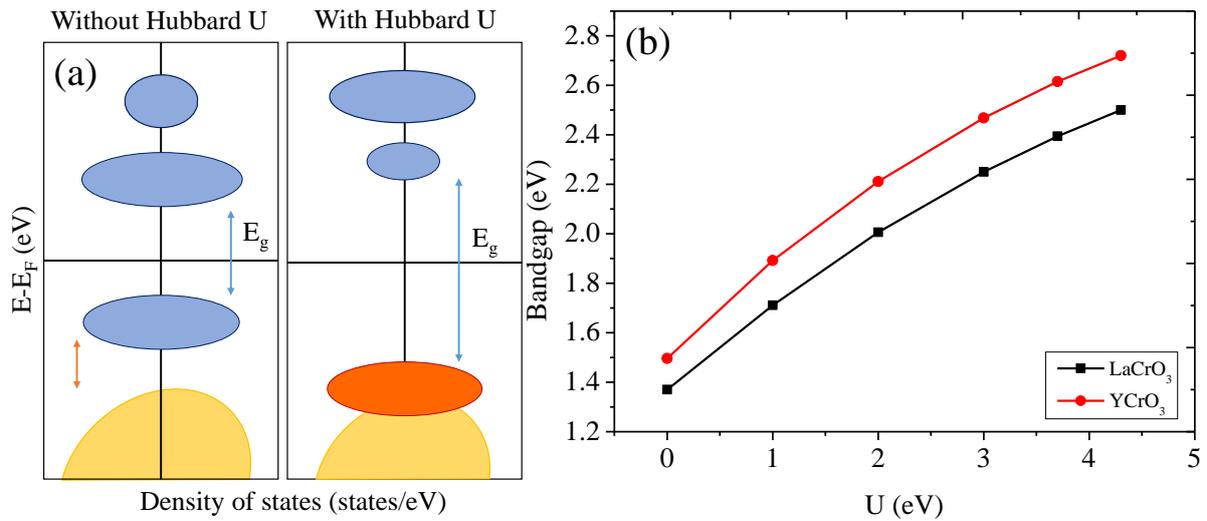

Figure 3: (a) Representation of how band gap is changing in ACrO$_3$ (A=La, Y), (b) Band gap changes with different Hubbard U in ACrO$_3$ (A=La, Y) system.

We further plotted molecular orbitals at the occupied level in Figure 4 to understand their contribution. These molecular orbitals represent the probability of finding electron $|\psi|^2 = \psi^*\psi$ and are not Wannier orbitals. Figure 4 (a1) shows the molecular orbital picture for LaCrO$_3$ without U where only axial oxygen (O11-O12) and 3d orbitals of Cr are contributing. As Hubbard energy is applied, contribution from non-axial oxygen's increases, as shown in Figure 4 (a2-a6), which is consistent with DOS, as shown Figure 2. Here, Cr-O1 and Cr-O2 bond lengths are smaller than other non-axial Cr-O bond lengths (Cr-O3, Cr-O4) as listed in Table 2. Therefore, O1 and O2 contribute first than O3, and O4 with increasing Hubbard energy because these are close to Cr. Both the 2p orbitals of axial and non-axial oxygen are contributing to the molecular orbital, as can be noticed from Figure 4 (a4) to 4(a6). The change in molecular orbitals for $CrO_6$ octahedra is shown in Figure 4 (c1-c4) with different U. YCrO$_3$ also exhibits nearly identical molecular orbital picture to that of LaCrO$_3$. The main difference is that in YCrO$_3$ both the axial oxygen (O11, O12), as well as partial contribution of non-axial 2p orbitals of oxygen (O1, O2), are contributing even without any Hubbard U applied, as can be seen in Figure 4(b1). Further, Figures 4(b2-b4) are showing similar change as explained in Figure 4(c3-c4). For LaCrO$_3$ the molecular orbitals following the nature from Figure 4(c1-c4) whereas for YCrO$_3$ it starts following from Figure 4(c2-c4).

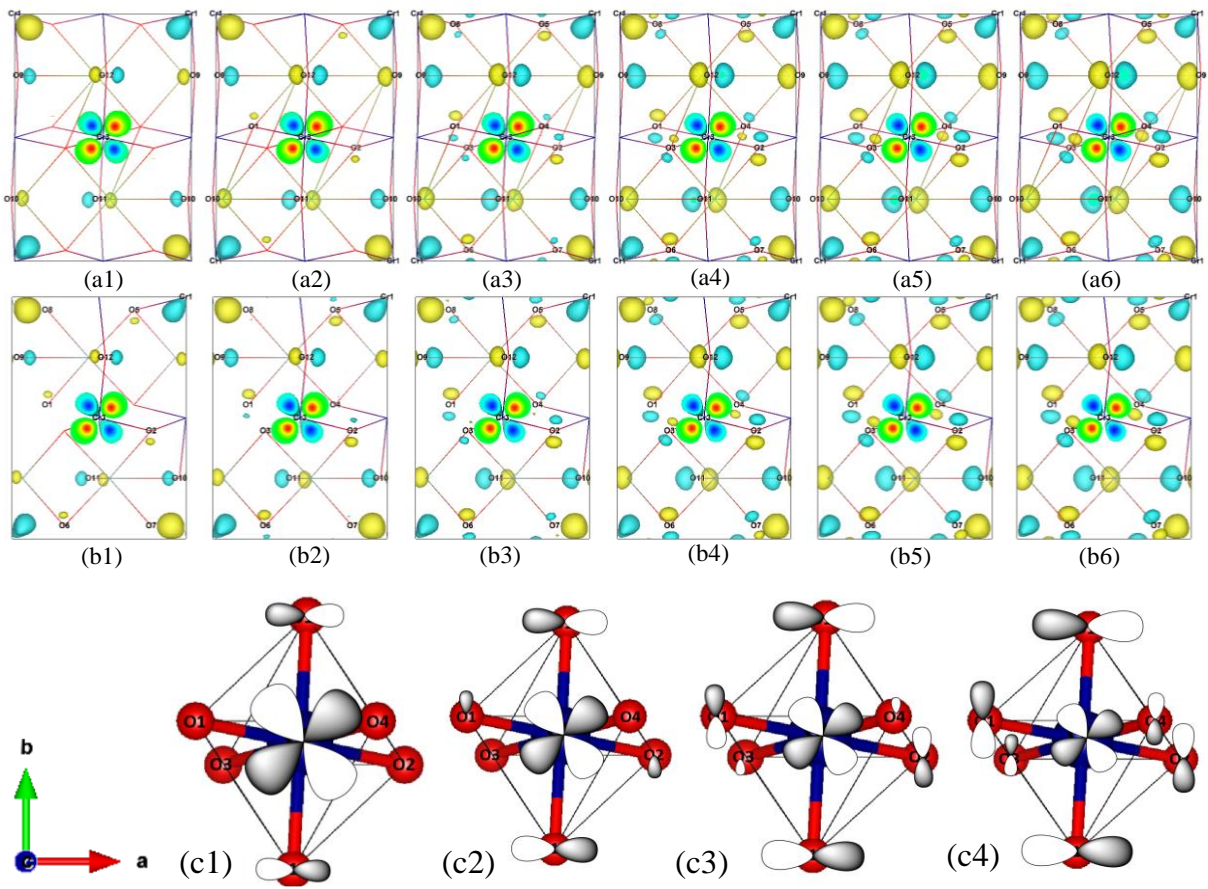

(a1) (a2) (a3) (a4) (a5) (a6)

(b1) (b2) (b3) (b4) (b5) (b6)

(c1) (c2) (c3) (c4)

Figure 4: Molecular orbital picture for (a1) LaCrO$_3$, (b1) YCrO$_3$ with U = 0 eV, (a2) LaCrO$_3$, (b2) YCrO$_3$ with U = 1 eV, (a3) LaCrO$_3$, (b3) YCrO$_3$ with U = 2 eV, (a4) LaCrO$_3$, (b4) YCrO$_3$ with U = 3 eV, (a5) LaCrO$_3$, (b5) YCrO$_3$ with U = 3.7 eV, (a5) LaCrO$_3$, (b5) YCrO$_3$ with U = 4.3 eV. Figure (c1-c4) shows the schematic diagram of molecular orbital picture change under the effect of Hubbard U in ACrO$_3$ (A=La, Y) system.

The noticed different contribution in molecular orbitals for LaCrO$_3$ and YCrO$_3$ is attributed to their different bond lengths and bond angles. In LaCrO$_3$, the molecular orbital picture changes with U linearly with the bond length. In this system, $Cr - O11(/12) < Cr - O1(/2) < Cr - O3(/4)$ (as listed in Table 2), therefore, only axial O (O11 and O12) are contributing when U =0 eV. As U increases the planer O1, O2 contribution increases followed by O3, O4. In contrast to LaCrO$_3$, YCrO$_3$ does not follow the same relation with bond length. For YCrO$_3$, $Cr - O1(/2) < Cr - O11(/12) < Cr - O3(/4)$., as noticed from Table 2. According to the bond length, it is expected that O1(/2) should contribute first followed by O11(/12) and O3(/4). But in YCrO$_3$, bond angle plays a major role than bond length. More interestingly, despite having different Cr-O bond length, molecular orbital picture for YCrO$_3$ follows the same trend as LaCrO$_3$. The molecular orbital picture suggests that the contribution of axial as well as non-axial oxygen increases with increasing U for both systems, and thus, substantiating the strong Cr-O hybridization. Here, the contribution of both oxygen's are increasing, while contribution of Cr is decreasing with increasing U.

Further, the partial density of states (PDOS) are plotted in Figure 5 ot understand the contribution of Cr and O orbitals in more detail. Figures 5 (a1) and (a2) show the PDOS for 3d orbitals of Cr with U =0 eV and with U = 4.3 eV, respectively for LaCrO$_3$. Figures 5 (b1) and (b2) show the PDOS for 3d orbitals of Cr with U =0 eV and with U = 4.3 eV, respectively for YCrO$_3$. Partial density of states (also known as projected density of states) provide the information about how each orbital is contributing. In ACrO$_3$ (A=La, Y), five Cr 3d-orbitals, i.e. ($d_{xy}, d_{yz}, d_{xz}, d_{x^2-y^2}, d_{z^2}$) show strong hybridization with three O 2p ($p_x, p_y, p_z$) orbitals. In ACrO$_3$, Cr is in +3 oxidation state, and $CrO_6$ octahedra distortion breaks the symmetry, causing splitting of d orbitals in higher and lower energy orbitals. The axial orbitals $e_g$ will have more energy as compared to the non-axial $t_{2g}$ orbitals in $CrO_6$ octahedra. Figures 5 (a1, a2) and (b1, b2), explains that Cr 3d orbitals split into $e_g$ ($d_{xz}, d_{x^2-y^2}$) and $t_{2g}$ ($d_{xy}$, d$_{zy}$, $d_{z^2}$) orbitals and the contribution of Cr is decreasing with increasing U, as discussed previously. It is consistent with Khuong et al.'s work [60]. Sabir et al. reported the case of Cr degeneracy due to compression of two Cr-O bond length, substantiating that the energy $d_{z^2}$ is higher than $d_{x^2-y^2}$ as observed in this work [58].

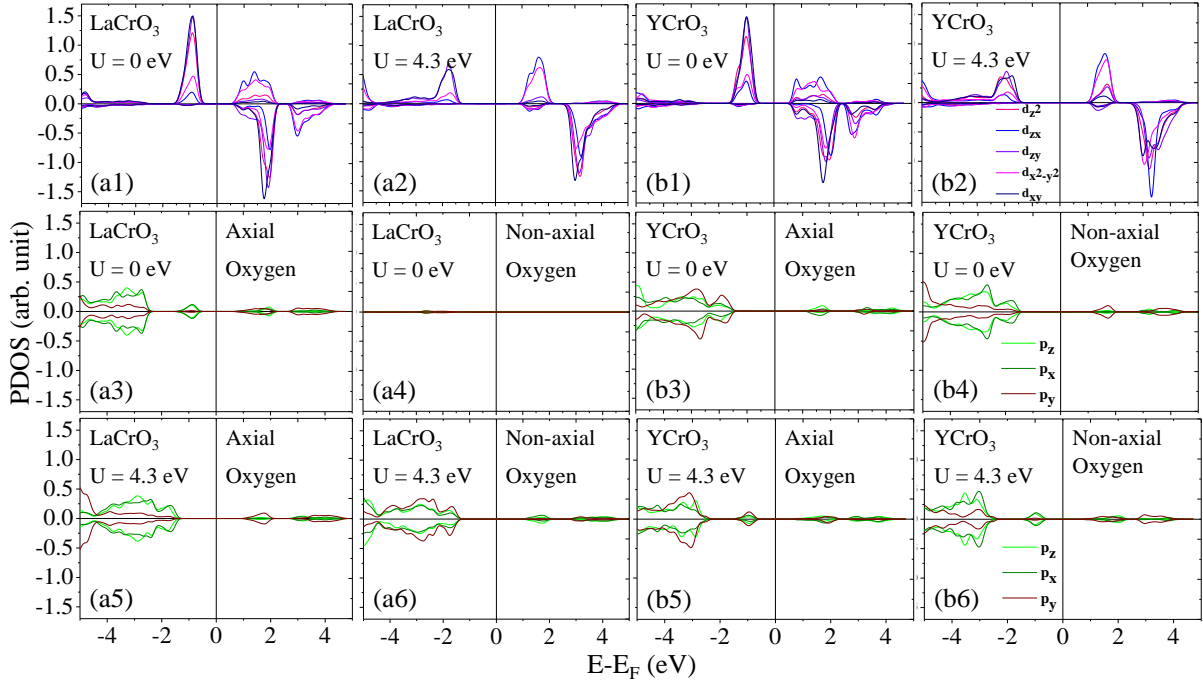

Figure 5: Partial Density of States of Cr 3d orbitals of ACrO3 with (a1, b1) U = 0 eV and (a2, b2) U = 4.3 eV. (Color line) shades of blue colors are showing the contribution of $t_{2g}$ orbitals and pink colors are showing the contribution of $e_g$ orbitals. (a3, b3) Oxygen contribution of axial oxygen and (a4, b4) non-axial oxygen for LaCrO3 with U=0 eV. Oxygen contribution of (a5, b5) axial oxygen and (a6, b6) non-axial oxygen for LaCrO3 with U=4.3 eV. Brown color represents the contribution of $p_y$ where green colors are represented contribution of $p_x$ and $p_z$ orbitals.

Figures 5 (a3-a4) and (b3-b4) show 2p axial and non-axial O orbitals PDOS at U = 0 eV for LaCrO3 and for YCrO3, respectively. Figures 5 (a5-a6) and (b5-b6) show 2p axial and non-axial O orbitals PDOS at U = 4.3 eV for LaCrO3 and for YCrO3, respectively. In the absence of U, only axial O are contributing in LaCrO3, i.e., no contribution from non-axial O. However, at U=4.3 eV, both axial and non-axial O orbitals are contributing equally. In contrast, for YCrO3 both axial and non-axial O are contributing near Fermi level even for U=0 eV, Figure 5 (b3-b6). For U= 0 eV, $p_x$ and $p_z$ orbitals are dominant in LaCrO3 whereas in YCrO3 $p_y$ orbitals are dominating only. For axial O, in LaCrO3 $p_x$ and $p_z$ are dominating whereas in YCrO3 for axial O $p_y$ is dominating irrespective of Hubbard energy. For non-axial O, $p_x$ and $p_z$ are dominating in YCrO3 irrespective of Hubbard energy. However, in LaCrO3, there is no contribution of O 2p orbitals in absence of Hubbard energy whereas $p_y$ orbitals are contributing in the presence of Hubbard energy. It is corroborated with the molecular orbital picture as shown in Figure 4.

**Magnetic Properties**

We also computed magnetic moments for $ACrO_3$ systems with U. The computed total magnetic moment is zero for different U as $ACrO_3$ systems are antiferromagnets. With increasing U, the magnetic moment increases linearly. The magnetic moments of $Cr^{+3}$ with S=3/2 are 2.90 µB and 2.94 µB with U = 0 for $LaCrO_3$ and $YCrO_3$ respectively, which get increased by 3.18 µB and 3.21 µB for U = 4.3 eV. The obtained results with Hubbard energy are close to experimental values 3.80 µB for $LaCrO_3$ and 4.5 µB for $YCrO_3$ [61,62]. Similar computational results were obtained by Tyagi et al. 2.8 µB with U=3 eV for bulk $LaCrO_3$ [63]. Magnetic moment of $YCrO_3$ is different from the theoretical and computational value due to its weak ferromagnetic nature. The effective magnetic moments are underestimated here because we have not include spin-orbital coupling in the calculation. Total Magnetic moment values of $ACrO_3$ with applied Hubbard U is listed in Supplementary S4.

Thus, the present work provides the microscopic origin of the onset of different physical properties for isostructural $LaCrO_3$ and $YCrO_3$ systems with different bond length and bond angles. The structural distortion drive the change in electronic properties of these systems and the bandgap is significantly affected by the Hubbard energy in $ACrO_3$ systems. Hubbard energy on Cr affects its energy level and Cr-O hybridization. The present findings substantiate that Hubbard energy plays an important role in Cr-O hybridization.

## 3. CONCLUSION

We demonstrated the microscopic origin of different physical and electronic properties of isostructural $ACrO_3$ (A=La, Y) systems. Rare-earth La and transition metal Y on A-site induce different structural distortion and it is more in $YCrO_3$ as compare to $LaCrO_3$. The present work demonstrates that the bond length plays major role in $LaCrO_3$, whereas bond angle plays crucial role in $YCrO_3$. It also affects the Cr-O hybridization differently in these systems, substantiated using molecular orbital pictures. Cr-O hybridization and energy bandgap are strongly affected by Hubbard energy U. and both increase with increasing U. The increase in band gap is attributed to the vanishing charge-transfer gap with increasing U. Thus, the contribution of U is important to understand $ACrO_3$ systems in more detail, which may affect the physical and electronic properties in such strongly correlated systems.

**Additional Information:**

VESTA package is used to draw the crystal structure of $ACrO_3$ [63]. Weighted-electronic band structures were plotted using gnuplot software.